%% file: paper-arxive.tex
\pdfoutput=1

\documentclass[conference]{IEEEtran}

\input{preamble}

\graphicspath{{figs/}}
\DeclareGraphicsExtensions{.pdf,.png,.jpg,.jpeg}

\def\BibTeX{{\rm B\kern-.05em{\sc i\kern-.025em b}\kern-.08em
    T\kern-.1667em\lower.7ex\hbox{E}\kern-.125emX}}

\begin{document}

\title{DRS-OSS: A Diff-Risk Scoring Tool for Continuous Integration Workflows}

\author{
\IEEEauthorblockN{Ali Sayedsalehi}
\IEEEauthorblockA{\textit{Concordia University}\\
Montreal, Quebec, Canada}
\and
\IEEEauthorblockN{Peter C. Rigby}
\IEEEauthorblockA{\textit{Concordia University}\\
Montreal, Quebec, Canada}
\and
\IEEEauthorblockN{Audris Mockus}
\IEEEauthorblockA{\textit{University of Tennessee, Knoxville}\\
Knoxville, Tennessee, USA}
}

\maketitle

\input{abstract}


\input{introduction}

\begin{figure*}[tb]
  \centering
  \includegraphics[width=\textwidth]{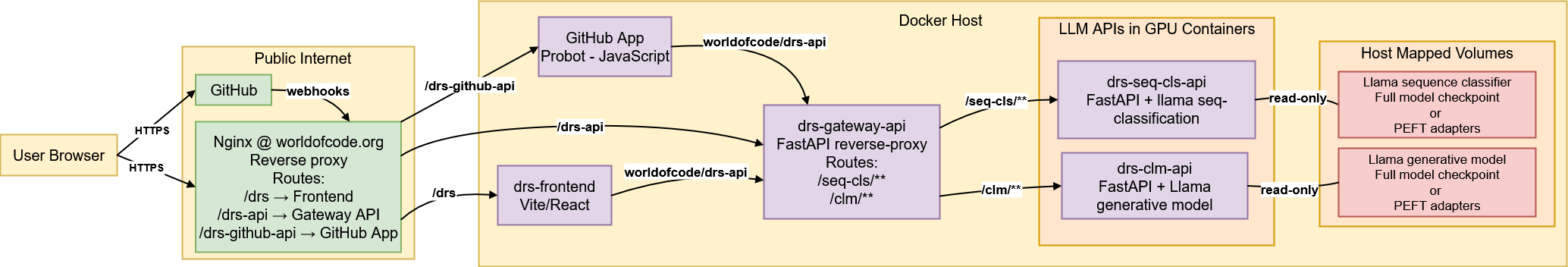}
  \caption{End-to-end DRS-OSS architecture.}
  \label{fig:arch}
\end{figure*}

\input{system-overview}

\input{model-api}

\input{web-ui}

\input{github-bot}

\begin{figure}[tbp]
  \centering
  \includegraphics[width=0.8\columnwidth]{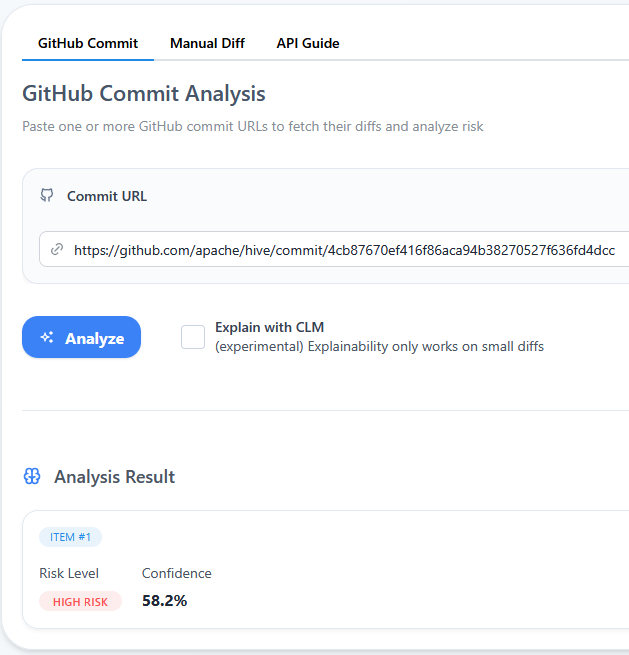}
  \vspace{-.1in}
  \caption{Demo result on the Hive out-of-memory fix commit, showing the derived label, risky or safe, and the confidence for the predicted label.}
  \label{fig:hive-demo}
  \vspace{-.1in}
\end{figure}

\begin{figure}[tbp]
  \centering
  \includegraphics[width=0.8\columnwidth]{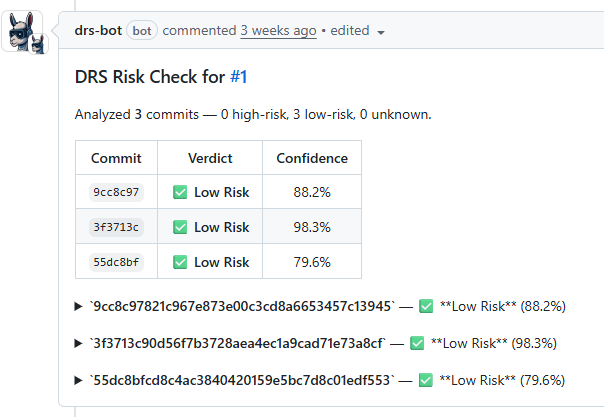}
  \vspace{-.1in}
  \caption{GitHub bot output for a pull request with three commits, showing the predicted label and confidence for each commit.}
  \label{fig:bot-comment}
  \vspace{-.1in}
\end{figure}

\input{deployment}

\input{tables/tool-usability}

\input{model-training}

\input{tool-customization}

\input{evaluation}

\input{threats-validity}

\input{conclusion}

\bibliographystyle{IEEEtran}
\bibliography{references}

\end{document}

%% file: preamble.tex
\usepackage{booktabs,tabularx,array,xspace,graphicx}
\usepackage{hyperref} 
\usepackage{fvextra} 
\usepackage{seqsplit}

\newcommand{\DRSOSS}{DRS-OSS\xspace}

\providecommand{\Description}[1]{}
\providecommand{\citep}[1]{\cite{#1}}
\providecommand{\citet}[1]{\cite{#1}}

\usepackage{fvextra} 

\DefineVerbatimEnvironment{drsverb}{Verbatim}{
  breaklines=true,
  breakanywhere=true,
  fontsize=\footnotesize
}

%% file: abstract.tex
\begin{abstract}
Software teams need change-risk scores that can guide CI decisions such
as review prioritization, test scheduling, and downstream
validation effort before risky changes are merged or released. However,
open-source teams often lack deployable tooling for surfacing these
risk signals in everyday CI workflows. We present \DRSOSS, an
open-source diff risk scoring tool for CI workflows.

\DRSOSS is designed as a deployable and customizable pipeline rather
than a standalone prediction model. It combines a REST API gateway,
containerized model services, a developer dashboard, GitHub integration,
and a replication package that lets users retrain or replace the
backend with other transformer models. The bundled workflow combines
commit messages, commit diffs, and change metrics in one risk
prediction pipeline. The default packaged backend uses a
Llama~3.1~8B sequence classifier configured for long diffs. Its
training recipe uses parameter-efficient tuning, quantization, CPU
offloading, and customization helper scripts so it can be adapted on modest hardware.

We compare \DRSOSS with similar tools and evaluate the bundled
classifier on ApacheJIT, where it reaches a ROC--AUC of 0.895 and
outperforms prior baselines. From a user-feedback perspective,
\DRSOSS\ has received interest from Uber, Duolingo, and Microsoft in
adapting the workflow to their own CI settings.

The full tool is released with source code, customization scripts, deployment artifacts~\cite{ReplicationPackage}, a public repository~\cite{GitRepo}, a live demo at worldofcode.org/drs, and a demonstration video available at https://www.youtube.com/watch?v=2FzeRRdNaco.
\end{abstract}

%% file: introduction.tex
\section{Introduction}\label{sec:intro}

Software teams need change-risk scores early enough to guide review and
CI decisions, not only after expensive validation work has already been
scheduled. Prior JIT-DP and deployment work suggests that these scores
are most useful when they operate at commit granularity, combine commit
content with established change metrics, and remain practical to retrain
and redeploy as repositories evolve~\citep{mockus2000jitdp,Kamei2013}.
Continuous risk probabilities also provide a ranked or thresholdable
signal for predictive test selection, code freeze gating, and review
routing~\cite{Machalica2018PTS,Kazemi2022riskReviewer}.
For open-source teams, this creates a systems need: low-latency diff
risk scoring that runs on modest infrastructure.

CommitGuru~\cite{Rosen2015CommitGuru} and
JITBot~\cite{Khanan2020JITBot} brought just-in-time defect prediction
(JIT-DP) into public commit-risk tooling~\citep{mockus2000jitdp,Kamei2013},
but they center traditional change metrics and repository-mined
features (size, entropy, files touched, etc.~\citep{Kamei2013}) rather than long-context diffs and commit messages. \DRSOSS\
extends this tooling space by jointly modeling commit messages, code
diffs, and change metrics, while also providing the full end-to-end,
deployable, and customizable system.

We present \DRSOSS, an open-source diff risk scoring tool and workflow
for CI settings. The system combines deployable model serving, inference API endpoints,
developer-facing interfaces, GitHub integration, and a documented
customization path so teams can consume risk probabilities through an
API or dashboard and still retrain or replace the backend as their
repositories evolve. The default classifier is a Llama-based model
trained on ApacheJIT~\cite{Keshavarz2022ApacheJIT}, while the
replication package also supports user-provided checkpoints or PEFT
adapters for other transformer families. The bundled backend uses
parameter-efficient tuning, 4-bit quantization, and DeepSpeed ZeRO-3
CPU offloading to support 22k-token contexts on a single 20\,GB GPU,
with quantized serving for practical long-diff inference.


%% file: system-overview.tex
\section{System Overview}\label{sec:system}
\noindent \DRSOSS packages event ingress, inference, policy handling, and result presentation into a single reproducible stack: a Gateway/API, model services, a Web UI, and a GitHub App. The system accepts either (i) PR/commit events from GitHub or (ii) manual submissions via the Web UI, and returns a label and confidence that can be consumed by reviewers, CI gates, or downstream automation (Fig.~\ref{fig:arch}).

\paragraph{Usage scenario.}
\begin{enumerate}
  \item \textbf{Event ingress.} A CI pipeline calls the inference API. A GitHub App subscribes to PR/commit webhooks. A user uploads a commit message+diff in the Web UI.
  \item \textbf{Gateway/API.} A FastAPI gateway routes requests to model services over REST.
  \item \textbf{Model services.} Each LLM microservice validates and normalizes payloads (repo, SHA, message, diff, metadata) and applies deployment-specific settings such as thresholds, max diff size, and timeouts.
  \item \textbf{Responses.} The Gateway combines model outputs with policy or business logic and returns label and confidence to the caller. For GitHub, a bot posts a PR comment; for the Web UI, it renders a result card.
\end{enumerate}

%% file: model-api.tex
\subsection{Model and API}\label{sec:model-api}

\begin{figure}[tbp]
  \centering
  \includegraphics[width=\columnwidth]{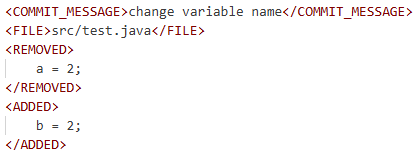}
  \vspace{-.2in}
  \caption{Commit structuring changes the unified diff into a simpler format that can be better understood by LLMs. It also reduces the number of input tokens in the sequence.}
  \Description{An XML text that shows tags for COMMIT_MESSAGE, FILE, REMOVED, and ADDED enclosing various commit and diffs lines.}
  \label{fig:struc-diff}
\end{figure}

The backend consists of a Gateway (FastAPI) and one or more Model services, each hosting a single LLM in its own Docker container. The Gateway exposes stable, documented endpoints and can be extended with database/authentication/business logic. Model services focus solely on inference and also handle input structuring of commits (Fig.~\ref{fig:struc-diff}). This separation keeps the external interface stable even when the deployed model changes. On a 32\,GB V100 GPU, end-to-end API response time ranges from under one second for small commits to about five seconds for large commits; this timing includes fetching the GitHub commit before scoring.

We provide two inference modes, both implemented using the Hugging Face pipeline API:
\begin{enumerate}
  \item \textbf{Sequence-classification head.} A fine-tuned Llama-3 model with a linear score head (replacing the LM head) produces risk probabilities as a label accompanied by a confidence score.
  \item \textbf{CLM-as-seq-cls.} For models without an exposed classifier head (e.g., closed-source models), we map the first generated token to \{0,1\} to obtain a label and confidence.
\end{enumerate}

Table~\ref{tab:endpoints} summarizes the inference API exposed by
the Gateway. The Web UI and GitHub bot are clients over these endpoints.

\input{tables/api-endpoints.tex}

Deployments can point the seq-cls service to (i) a full checkpoint or (ii) a PEFT adapter merged at load time. Users mount their model artifacts via a host-mapped docker volume and configure paths through environment variables. This allows teams to swap the bundled classifier for their own model without rebuilding the surrounding toolchain.

Each LLM service can optionally use 4-bit weight quantization to accommodate long diffs and increase throughput. The models can be configured with mixed-precision or full-precision, sequence length limits, and quantization control.

%% file: tables/api-endpoints.tex
\begin{table}[t]
\caption{Inference API endpoints.}
  \label{tab:endpoints}
  \centering
  \scriptsize
  \begin{tabularx}{\columnwidth}{>{\ttfamily\raggedright\arraybackslash}p{0.47\columnwidth}X}
    \toprule
    \textbf{Method \& Path} & \textbf{Purpose} \\
    \midrule
    GET /health & Health check \\
    POST /seq-cls/predict & Commit message + unified diff $\rightarrow$ label, confidence \\
    POST /seq-cls/predict\_batch & Batch input $\rightarrow$ label, confidence per commit \\
    POST /seq-cls/predict\_by\_sha & GitHub Repo + commit SHA $\rightarrow$ label, confidence \\
    \bottomrule
  \end{tabularx}
  \vspace{-.1in}
\end{table}

%% file: web-ui.tex
\subsection{Web UI}\label{sec:webui}
Two workflows are supported via the Web interface: \emph{GitHub Commit} (paste a commit URL; the Gateway resolves repo/SHA and diff) and \emph{Manual Analysis} (paste a commit message and a unified diff). Both return a label and confidence score. This makes it possible to inspect a single change even when the GitHub App is not installed. The frontend has its own Express server and uses Vite, React, and Mantine as its tech stack.

We illustrate Web UI usage on a real change from \texttt{apache/hive} that fixes an out-of-memory error. Bug-fix changes, especially those touching resource management and error-handling paths, are known to be error-prone due to their complexity and cross-cutting effects. Fig.~\ref{fig:hive-demo} highlights how risk scoring is displayed in the Web UI.


%% file: github-bot.tex
\subsection{GitHub Bot}\label{sec:bot}
The GitHub App integrates with repository workflows and supports both automatic and on-demand operation.
The app is installed on selected repositories with read access to
commit messages and code, and write access to PR comments/status
checks. On pull request (opened/synchronize) or
push, the app fetches the diff and commit message, calls
the Gateway’s prediction endpoints, and posts the result. Reviewers
can also request an explicit prediction by commenting \texttt{/drs}
on a PR. The bot posts a compact card: commit/PR identifiers,
derived label (risky/safe), and confidence score.

%% file: deployment.tex
\subsection{Deployment}\label{sec:deploy}
\DRSOSS is designed to be reproducible and deployable on a team's own
infrastructure. Each LLM model
runs in its own Docker container to simplify scaling and upgrades. A
Docker Compose configuration brings up the Gateway, Model services, the
frontend, and the GitHub App. Model services utilize CPU offloading and
eagerly load weights on startup.

An Nginx reverse proxy terminates TLS and routes
\texttt{/drs-api} to the Gateway and \texttt{/drs} to the Web UI. A
single environment file sets model IDs/paths, dtype, maximum
sequence length, and quantization flags. This packaging lets users
evaluate the bundled model as-is or replace it with their own
checkpoint or adapter without changing the surrounding workflow
components.

The replication package also includes prepared scripts for dataset
construction and modification, as well as for retraining or fine-tuning
a compatible transformer with the same memory-efficient recipe used by
the bundled model. Once trained, the resulting checkpoint or adapter can
be redeployed by mounting it into the model service's Docker volume and
updating the configured path.

%% file: tables/tool-usability.tex
\begin{table*}[t]
  \caption{Operational comparison of \DRSOSS\ with public commit-risk tooling.}
  \label{tab:tool-usability}
  \centering
  \small
  \renewcommand{\arraystretch}{1.2}
  \begin{tabularx}{\textwidth}{l
      >{\raggedright\arraybackslash}X
      >{\raggedright\arraybackslash}X
      >{\raggedright\arraybackslash}X}
    \toprule
    \textbf{Tool} & \textbf{Input modeling} & \textbf{Interface} & \textbf{Classifier Model} \\
    \midrule
    CommitGuru~\cite{Rosen2015CommitGuru} & Change metrics & Web UI & Logistic Regression\\
    JITBot~\cite{Khanan2020JITBot} & Change metrics & GitHub bot & Random Forest \\
    \textbf{\DRSOSS} & Long-context diff, commit message, change metrics, input structuring & REST API, Web UI, GitHub bot & Customizable  (Default: Llama with PEFT, quantization, CPU offloading) \\
    \bottomrule
  \end{tabularx}
\vspace{-.1in}
\end{table*}

%% file: model-training.tex
\section{Classifier Model Training Method}\label{sec:training}

This section documents the training method used for the default classifier model artifact shipped with \DRSOSS. The goal is to make the demo reproducible and to show how the packaged classifier can be retrained or replaced; the tool itself is not tied to this specific model. We choose Llama as the bundled base model because it has open weights, strong joint pretraining on natural language and code, and can be deployed cleanly inside the Hugging Face-based model services used by \DRSOSS.

For the default packaged model, we fine-tune Llama~3.1~8B with a
classifier head. \DRSOSS\ also supports the
alternative LM-head setup, where the model emits label tokens such as 0
or 1, which is useful when the deployed backend exposes only
generative-text outputs, such as for closed-source models. The classifier-head setup is more efficient in
both training and inference because the model predicts one risk score
directly rather than scoring the full output vocabulary. We
experimented with both setups and observed very similar predictive
performance, so we use the classifier-head version as the default
model. In this classifier-head setup, the transformer first encodes the
full commit into contextualized token representations. We then use the
hidden state of the last non-pad token as a compact summary of the
entire input and feed it to a small linear layer that produces one
score, which a sigmoid converts into the final defect-risk probability.

We use ApacheJIT~\citep{Keshavarz2022ApacheJIT}, a large public JIT-DP benchmark spanning 14 Apache projects with 106,674 commits (28,239 bug-inducing and 78,435 clean commits), to train and validate
the default classifier. We
use an 80/10/10 chronological split (train/validation/test) to avoid
temporal leakage, ensuring that the model is trained only on past
commits and never exposed to commits from the future.

The inputs to the model are structured with special delimiters to separate commit message, code diff, and change metrics~\citep{Keshavarz2022ApacheJIT}. We add these delimiters as new tokens to the model during fine-tuning so that it can learn their structural role. These tokens are shown in Fig.~\ref{fig:struc-diff}. To expose scalar change metadata alongside natural language and diff text, we serialize metrics as bucketed tokens at the start of the sequence. The special tokens are added to the tokenizer vocabulary and the embedding matrix is correspondingly resized. Placing metrics first gives the encoder early access to coarse change signals while still allowing the model to attend across message and diff content. After normalization, bucketing, and structuring, the metrics section of each training sample has the following format:

\begin{verbatim}
[num_lines_added:] [LOW]
[num_lines_deleted:] [HIGH]
[num_files_touched:] [VERY_LOW]
[num_directories_touched:] [VERY_HIGH]
[num_subsystems_touched:] [VERY_LOW]
[change_entropy:] [MEDIUM]
[num_developers_touched_files:] [HIGH]
[time_from_last_change:] [VERY_LOW]
[num_changes_in_files:] [VERY_LOW]
[author_experience:] [LOW]
[author_recent_experience:] [MEDIUM]
[author_subsystem_experience:] [VERY_HIGH]
\end{verbatim}

Class imbalance is addressed with undersampling of the majority class
using a 0.7 sampling ratio during batch construction. Resampling is
applied only to the train split to keep evaluation and test splits
unchanged.

Long diffs are a practical constraint, so we use 4-bit quantization,
LoRA adapters on attention and MLP modules, and a maximum sequence
length of 22{,}000 tokens. Training uses DeepSpeed ZeRO-3 with CPU
offloading, gradient accumulation, and mixed precision; on a single
20\,GB A100, the 3-epoch run completes in roughly two weeks. These
choices keep the bundled model reproducible on modest hardware and
compatible with the containerized inference services described earlier.

%% file: tool-customization.tex
\section{Tool Customization}\label{sec:customization}

The replication package is designed not only to reproduce the bundled
Llama-based classifier but also to help users retrain \DRSOSS with
other transformer backbones. The main training entry point in
\texttt{llama/train.py}, reads YAML configurations and local
model snapshots, supports both sequence-classification heads and
CLM-as-seq-cls training, and detects both encoder-style families
(e.g., BERT, RoBERTa, and CodeBERT) and decoder-style families
(e.g., Llama and StarCoder). There are helper scripts to attach a classifier
head to a causal LM or merge LoRA adapters into a standalone
checkpoint. For sequence-classification backbones, the package also
supports configurable pooling of token representations, including
``last``, ``mean``, and ``max`` pooling, which lets users
adapt the classifier head behavior to different model families and data
conditions. Briefly, ``last`` uses the final non-padding token
state, ``mean`` averages non-padding token states across the
sequence, and ``max`` keeps the strongest activation per hidden
dimension; this gives users a simple way to choose whether the classifier
should emphasize the tail of the sequence, distribute evidence across a
long diff, or focus on sparse high-signal tokens.

The package also exposes practical data handling through the main
preprocessing entry point,
\texttt{\seqsplit{data\_extraction/data\_preparation.py}}.
That script produces multiple dataset variants for JIT defect
prediction data, sorts examples
chronologically, converts raw Git or Mercurial diffs into
XML-like structured format (Fig.~\ref{fig:struc-diff}), bucketizes scalar
metrics into discrete prompt tokens, and can emit either
sequence-classification or CLM-style inputs with optional metadata
prefixes. Supporting utilities clean noisy commit-message prefixes and
handle renames and binary-file cases during diff structuring. During
training, the loader can stage JSONL datasets on Slurm node-local
storage, preserve chronological train/validation/final-test splits, and
handle class imbalance through oversampling, undersampling, weighted
loss, or focal loss. Each training run
stores its effective configuration, checkpoints, logs, and held-out
predictions in a timestamped directory.

Efficiency is treated as a first-class concern in the package. The
scripts support 4-bit quantization, LoRA-based adapter tuning,
gradient checkpointing, BF16/FP16 mixed precision, optional
FlashAttention~2, and Hugging Face accelerate configurations for
DeepSpeed ZeRO-3 or FSDP. The code also infers LoRA target modules
automatically across encoder and decoder architectures and, when new
structural tokens are added, can restrict PEFT updates to the newly
introduced embedding rows. New input and output-side control tokens are
configured through a single YAML field, \texttt{new\_tokens}, which can
list any number of additional delimiters, metric buckets, or task
markers. The trainer adds only missing tokens to the tokenizer, resizes
the embedding matrix once, and then updates only the new rows when
possible. This makes it practical for users to extend the structured
representation with extra signals such as test metadata, platform tags,
or new defect-prediction metrics without rewriting the training code or
the surrounding tool interfaces.

For users who want a practical starting point, the replication package
includes both Docker and Slurm helper scripts for all customization tasks. The scripts prepare
a CUDA-based environment, configure cluster
modules, virtual environments, and runtime flags for accelerated Hugging Face training. Together, these
artifacts show that users can swap in a new dataset, a different
backbone, or a cluster-specific execution recipe while preserving the same external tool interface.

%% file: evaluation.tex
\section{Tool Evaluation}\label{sec:evaluation}

This section summarizes both the practical usability of \DRSOSS\ as a
tool and the predictive behavior of its bundled default classifier.

\subsection{Operational Utility}
Table~\ref{tab:tool-usability} compares \DRSOSS\ with public
commit-risk tools. The comparison highlights the design goal of
\DRSOSS: to combine deployable tooling interfaces with long-context
diff modeling and practical efficiency for OSS teams. Relative to
CommitGuru~\cite{Rosen2015CommitGuru} and
JITBot~\cite{Khanan2020JITBot}, \DRSOSS\ moves beyond metric-centered
commit analytics by jointly modeling commit messages, code diffs, and
change metrics in one pipeline. It also offers a broader interface
surface through a REST API, Web UI, and GitHub bot, while the
replication package documents how to retrain or swap the backend model
without changing those interfaces.

\subsection{Performance of the Default Classifier}
Because \DRSOSS\ ships with a default classifier for the demo, we
report an offline evaluation to document the behavior of that bundled
model.

Table~\ref{tab:main-metrics} compares the bundled \DRSOSS\
classifier against prior JIT-DP models reported on ApacheJIT under the
same dataset conditions and benchmark settings.
The bundled model achieves the highest ROC--AUC and F1 on ApacheJIT
among the compared JIT-DP baselines.

\input{tables/metrics.tex}

Many teams act only on the highest-risk slice of changes, so
Recall@Top-$k\%$ is a natural operational metric: it measures the
fraction of truly risky changes captured if only the riskiest $k\%$ of
diffs are quarantined or sent for extra scrutiny. Table~\ref{tab:recall-topk}
shows that sending only the riskiest 30\% of commits for extra scrutiny
captures 86.4\% of risky commits in the test set, giving users a
practical starting point for local gates and review-prioritization
policies.

\input{tables/recall-topk.tex}

\subsection{User Feedback}\label{sec:feedback}
The tool has attracted interest from Uber, Duolingo, and Microsoft. We had discussions with these companies on how the workflow could be inspected, reproduced, or adapted
for their own repositories. The main feedback from these companies
was that the system should also support third-party LLM APIs, e.g. OpenAI, and RAG workflows,
which we plan to address in future iterations. This feedback reinforces that \DRSOSS\ is a practical
starting point for teams that want deployable diff risk scoring with a
customizable backend.

%% file: tables/metrics.tex
\begin{table}[t]
  \caption{ApacheJIT test metrics for \DRSOSS\ and prior JIT-DP models at their reported operating points (higher is better).}
  \label{tab:main-metrics}
  \centering
  \footnotesize
  \begin{tabular}{lccccc}
    \toprule
    \textbf{Model} & \textbf{F1} & \textbf{ROC-AUC} & \textbf{Prec.} & \textbf{Rec.} & \textbf{Acc.} \\
    \midrule
    JITGNN~\cite{keshavarz2024jitgnn} & 0.5 & 0.81 & 0.37 & 0.78 & - \\
    NeuroJIT~\cite{lee2024NeuroJIT} & 0.56 & 0.8 & - & - & - \\
    CodeReviewer~\cite{AbuTalib2024} & 0.622 & 0.815 & 0.497 & 0.83 & 0.806 \\
    \textbf{\DRSOSS}  & \textbf{0.641} & \textbf{0.895} & 0.491 & 0.84 & 0.802 \\
    \bottomrule
  \end{tabular}
\end{table}

%% file: tables/recall-topk.tex
\begin{table}[t]
  \caption{Recall@Top-$k\%$ for the bundled \DRSOSS\ classifier (higher is better).}
  \label{tab:recall-topk}
\centering
  \small
  \begin{tabular}{lccc}
    \toprule
    \textbf{recall@5\%} & \textbf{recall@10\%} & \textbf{recall@30\%} \\
    \midrule
    0.315 & 0.503 & 0.864 \\
    \bottomrule
  \end{tabular}
\vspace{-.1in}
\end{table}

%% file: threats-validity.tex
\section{Threats to Validity}\label{sec:threats}

One threat to the offline evaluation is pretraining contamination:
Llama may have seen Apache code during pretraining, so some ApacheJIT examples may not be entirely novel at evaluation time. We
therefore treat the ApacheJIT results as a practical benchmark result,
not a contamination-free estimate of generalization.

%% file: conclusion.tex
\section{Conclusion}\label{sec:conclusion}
This paper introduced \DRSOSS, an open-source tool that packages diff
risk scoring into a deployable workflow for OSS projects. By combining a
reproducible serving stack, practical developer-facing interfaces, and
a documented customization path, the system surfaces risk scores in
everyday CI and code-review settings while remaining adaptable to a
team's own data, models, and infrastructure.